\documentclass[preprintnumbers,preprint,amsmath,amssymb]{revtex4}

\usepackage{graphicx}
\usepackage{dcolumn}
\usepackage{bm}

\begin{document}


\title{Practical quantum key distribution over 60 hours at an optical fiber distance of 20km using weak and vacuum decoy pulses for enhanced security}

\author{J. F. Dynes}
\email{james.dynes@crl.toshiba.co.uk} 
\author{Z. L. Yuan}
\author{A. W. Sharpe}
\author{A. J. Shields}
\affiliation{Toshiba Research Europe Limited, Cambridge Research Laboratory, 260 Cambridge Science Park, Cambridge, CB4 0WE, UK}

\date{\today}

\begin{abstract}
Experimental one-way decoy pulse quantum key distribution running continuously for 60 hours is demonstrated over a fiber distance of 20km. We employ a decoy protocol which involves one weak decoy pulse and a vacuum pulse. The obtained secret key rate is on average over 10kbps. This is the highest rate reported using this decoy protocol over this fiber distance and duration.  
\end{abstract}

\keywords{Quantum optics; Photon statistics; Fiber optics and optical communications}
\maketitle
\section{Introduction}
Quantum key distribution (QKD) involves the secure communication between two remote parties, conventionally named Alice (sender) and Bob (receiver), where the security of the keys is determined by the laws of quantum mechanics rather than the use of strong, one-way  mathematical functions to encrypt the keys \cite{gisin2002}. Where security is of paramount importance, QKD naturally would be the technique of choice for secret key distribution owing to its unconditionally secure nature. Although since the original proposal \cite{bennett1984}, there has been a considerable amount of work on QKD, beginning with the first experimental demonstration in 1992 \cite{bennett1992}, reliable, compact systems with tolerably high enough bit rates compatible with existing telecoms fiber technology are only now starting to emerge \cite{stucki2002,gobby2004}. Ideally the QKD setup should be arranged employing a true single photon source to guarantee immunity against the so-called pulse number splitting attacks (PNS)\cite{wang2005,brassard2000} from a potential eavesdropper (Eve). Presently there are a lack of deterministic, reliable and useful single photon sources. Therefore most implementations rely on heavily attenuated lasers as a photon source which emit photon pulses with a Poissonian number distribution. The PNS attack in such a configuration would consist of blocking any true single photons in the quantum channel, removing part of the multi-photon pulses and transmitting the remaining portion to Bob via a lossless channel. In Eve's best case scenario, Bob's detection rate is maintained while Bob is oblivious to Eve's presence. Eve can then determine all or part of the key \cite{brassard2000}.

A recent proposal \cite{hwang2003} circumvents the PNS attack using additional (decoy) pulses sent by Alice. These pulses in general have different pulses intensities compared to the signal pulses and are interleaved randomly with the signal pulses. Bob measures the transmittances of the quantum channel for these different pulse intensities and can then infer tighter bounds on the final secure key by the ability to detect PNS attacks. 
\begin{figure}[htb!]
\centerline{\scalebox{0.4}{\rotatebox{270}{\includegraphics*[15mm,15mm][205mm,265mm]{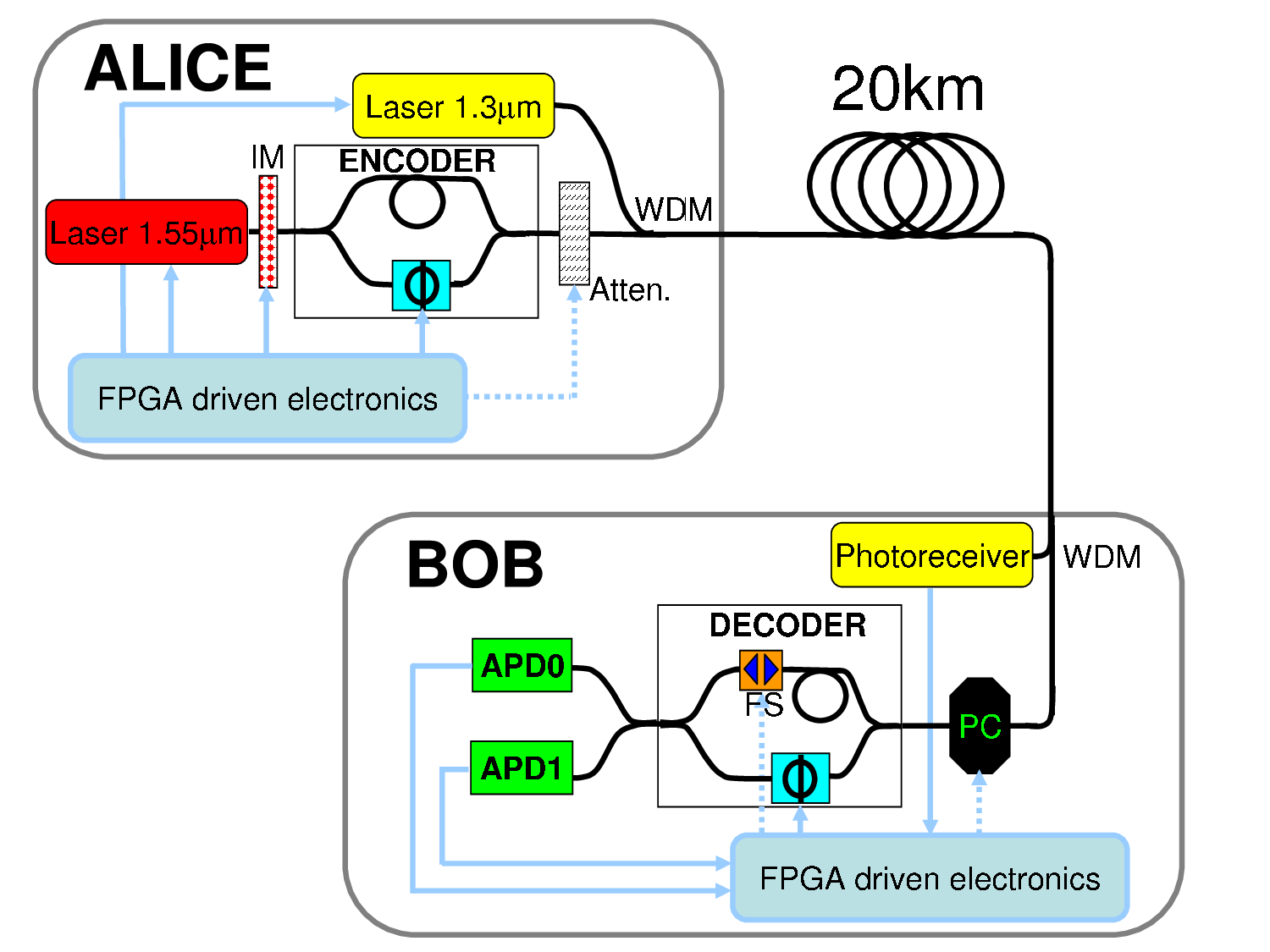}}}}
\caption{\label{fig:qkd_setup} Schematic of the optical layout of the one-way quantum key distribution system. The system employs BB84 phase encoding and the weak + vacuum decoy protocol. Attn: attenuator, IM: intensity modulator,  PC: polarization controller, WDM: wavelength division multiplexer, FS: fiber stretcher. The QKD optics are driven by field programmable gate arrays (FPGAs) electronics. Fast (MHz) electronic pulse routes: solid arrows; slow (Hz) electronic pulse routes: dotted arrows.}
\end{figure}
Under such conditions Eve is powerless to attack the channel using the PNS. Recently a proof of the decoy pulse protocol has been displayed (GLLP)\cite{gottesman2004} which also includes realistic (if pessimistic) experimental assumptions; Bob's detectors can be manipulated by Eve and the detectors have finite detection efficiency. A recent experimental bi-directional decoy pulse system has been implemented \cite{zhao2006}  but (as pointed out by Ref. \cite{zhao2006}) a drawback of bi-directional systems are the pulse intensities can indeed be tampered with by Eve unknown to Alice and Bob, thus compromising security. Very recently we showed an extremely promising one-way QKD system employing a single decoy pulse \cite{yuan2007}. This system displayed a high key rate of over 5kbps with 25km of optical fiber. For a non-decoy system using much quieter detectors a secure bit rate of just 43bps was achieved over the same length of fiber \cite{gobby2004b}. This illustrates the power of the decoy pulse method in providing a much more stringent bound on the security of the final key. Even tighter bounds can be achieved using more than one decoy pulse \cite{wang2005}. Theory shows that it is usually enough to limit the number of decoy pulses to two decoy pulses only. The optimal case for the two decoy pulses is for one decoy of which is a lower intensity $\nu$ than the signal pulse $\mu$ and another which has an intensity close to zero or ``vacuum-like" \cite{ma2005}. The lower bound on the single photon gain, $Q_{1}^{L}$ is then given by three transmittances, signal, decoy and vacuum respectively: $Q_\mu$, $Q_\nu$ and $Y_{0}$ \cite{ma2005,zhao2006}: 
\begin{equation}
Q_{1}^{L}=\frac{\mu^2e^{-\mu}}{\mu\nu - \nu^2}\{Q_{\nu}^{L}e^\nu - Q_{\nu}e^\mu\frac{\nu^2}{\mu^2} -Y_{0}^{U}\frac{\mu^2 - \nu^2}{\mu^2}\}
\label{eq:single_photon_gain}
\end{equation}
where $Q_{\nu}^{L}$ ($Y_{0}^{U}$) are the lower (upper) bounds on the decoy pulse and vacuum transmittances respectively , estimated conservatively as ten standard deviations of $Q_{\nu}$ ($Y_{0}$) from the measured value ensuring a confidence interval of $1-1.5\times10^{-23}$. The bit errors are assumed to derive from the subset of single photon pulses and the upper bound for the single photon error rate can be written as:
\begin{equation}
\epsilon_{1}^{U}=\frac{\epsilon_{\mu}Q_{\mu}}{Q_{1}^{L}} - \frac{Y_{0}^{L}e^{-\mu}}{2Q_{1}^{L}}
\label{eqn:error_rate}
\end{equation}
where $\epsilon_{\mu}$ is the signal error rate and $Y_{0}^{L}$ is the lower bound on the vacuum transmittance (estimated as as ten standard deviations of $Y_{0}$ from the measured value). 
The lower bound on the final secure key rate, $R^{L}$ can then be determined by the following expression:
\begin{equation}
R\geq R^{L}=qN_{\mu}\{-Q_{\mu}f(\epsilon_{\mu})H_{2}(\epsilon_{\mu}) + Q_{1}^{L}(1-H_{2}(\epsilon_{1}^{U}))\}/t
\label{eq:key_rate}
\end{equation}
where $q=0.5$ for the BB84 protocol, $N_{\mu}$ is the total number of signal pulses sent by Alice, $f(x)$ is the bi-directional error correction efficiency above the Shannon limit is estimated to be $f(\epsilon_{\mu})\sim1.10$, $H_{2} = -x$log$_{2}(x) - (1-x)$log$_{2}(1-x)$ is the binary entropy function  and the time $t$ is the duration of the key session. The first term in eq. (\ref{eq:key_rate}) corresponds to error correction; the second term corresponds to the single photon gain modified by privacy amplification ($H_{2}(\epsilon_{1}^{U})$). 
Although the weak + vacuum decoy protocol is predicted to have an improved performance over the single decoy pulse protocol, recent implementations \cite{peng2007,rosenberg2007} show relatively low key generation rates and are limited to local synchronization.  

Here we present a one-way weak + vacuum decoy pulse system which has a relatively high secure continuous key rate of more than 10kbps over 20km of fiber length. The key generation rate is stable over the rate time period of 60 hours and the system requires no manual alignment for set-up or during operation. 
\begin{figure}[htb!]
\centerline{\scalebox{1.25}{\rotatebox{0}{\includegraphics*[5mm,5mm][125mm,56mm]{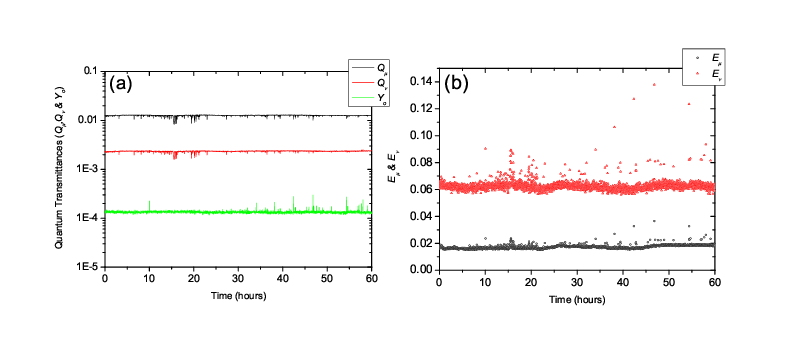}}}}
\caption{\label{fig:transmittances_and_qbers}Experimentally measured data for the 60 hours experiment. (a)Transmittances of the signal, decoy and vacuum. (b) The quantum bit error rates of the the signal ($E_{\mu}$) and the non-zero decoy ($E_{\nu})$ as a function of time. } \end{figure}
\begin{figure}[htb!]
\centerline{\scalebox{1.25}{\rotatebox{0}{\includegraphics*[5mm,5mm][108mm,80mm]{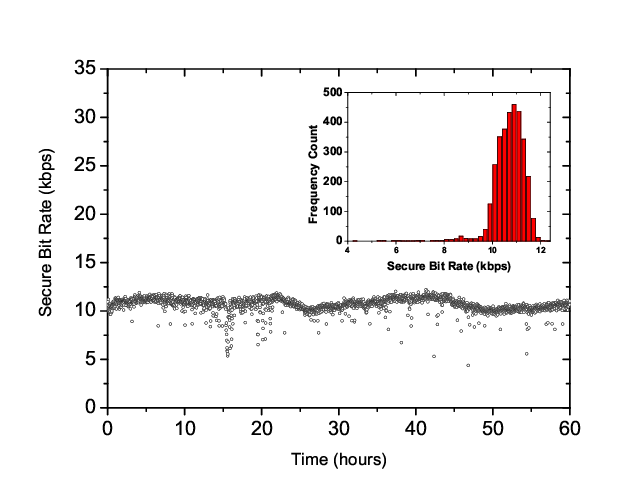}}}}
\caption{\label{fig:rates}The final secure bit rate as a function of time. The extremely long term drift in the key rate is attributed to long term day to day temperature drift in the laboratory. Inset: distribution of the secure bit rates for the keys. } \end{figure}
\section{QKD experimental setup}
We use a one-way fiber optic QKD system with phase encoding as shown in Fig. \ref{fig:qkd_setup}. Two Mach-Zehnder phase encoding interferometers are employed for the phase encoding (Alice; sender) and decoding (Bob; receiver). Alice and Bob have a 20km fiber spool linking them through which the signal ($\lambda=1.55\mu$m) and clock ($\lambda=1.3\mu$m) optical pulses are transmitted. The $1.3\mu$m clock pulse duration is $5$ns with a peak intensity of $\sim5\mu$W; average intensity is $\sim200$nW at Alice. The clock pulse over the entire transmission distance does not overlap the ($1.55\mu$m) signal pulses. The signal laser is a distributed feedback type and emits a fixed intensity train of pulses at a repetition rate of 7.143MHz. An intensity modulator is used to produce signal and decoy pulses of differing intensities. The vacuum decoy pulse is produced by omitting trigger pulses to the signal laser. All signal, decoy and vacuum pulses are produced at random times and can have relative occurance probabilities assigned to them. The signal and decoy pulses are attenuated strongly to the single photon level after which a much stronger clock pulse is wavelength division multiplexed with them to provide synchronization between Alice and Bob's electronics.  Customized electronics based on FPGAs were developed in house to drive the QKD optics. An active stabilization technique is employed to ensure continuously running operation.  Bob's detectors are two single photon InGaAs avalanche photodiodes (APDs) cooled to approximately $-30^{o}$C and characterized to have negligible afterpulsing \cite{ribordy1998}. . Their combined dark count rates are $\sim1.4\times10^{-4}$. Bob's detector efficiency $\sim 10$\% and loss $\sim 2.5$dB gives rise to an overall efficiency of $5.62\times10^{-2}$.
The weak + vacuum including BB84 protocol was implemented. Numerical simulation to maximize the secure bit rate was performed to yield the optimal intensities of the signal and decoy pulses as $\mu =0.55 $ and $\nu = 0.098$. Numerical simulation also provided the optimal probabilities of pulses: signal $N_{\mu} = 0.93$, decoy pulse $N_{\nu}=0.062$ and vacuum pulse $N_{0} = 0.016$. The session length for each QKD key is selected as $3\times10^{6}$ bits which corresponds to roughly $6\times10^{6}$ detection events by Bob. A total of 3262 sessions were distributed with an average individual session time of $\sim 71$ seconds. 

\section{Results}
Figure \ref{fig:transmittances_and_qbers}(a) shows the experimentally measured transmittances of the signal, decoy and vacuum pulses. The values are stable and constant over the duration of the experiment and agree well with the theoretically predicted transmittance (per pulse): $Q_{\mu}=0.01270\pm0.00078$, $Q_{\nu}=0.00234\pm0.00014$ where the errors are two standard deviations; for the vacuum transmittance (per pulse), $Y_{0} = 1.34\pm 0.20\times10^{-4}$ in good agreement with the measured dark count value of $\sim1.4\times10^{-4}$  measured prior to the experiment. A small (simultaneous) proportion of fluctuations in $Q_{\mu}$ and $Q_{\nu}$ are observed and are attributed to polarization and/or fiber stretcher resets during which photons were temporarily not counted. These obtained transmittances indicates the various optical states had been well prepared and detected. The associated quantum bit error rates of both the signal ($E_{\mu}$) and the non-zero decoy ($E_{\nu}$) are plotted in Fig. \ref{fig:transmittances_and_qbers}(b). They are fairly stable and constant. The final secure bit rate is displayed in Fig. \ref{fig:rates}. A secure bit rate of $>10$kbps is observed. The long term drift in the key rate is attributed to long term temperature drift from day through to night (the period is roughly 24 hours) in the laboratory affecting the overall temperature of the 20km fiber spool. In a real world environment the fiber is usually located around 1 meter underground leading to very stable fiber temperatures. This would eliminate this long term drift observed here. 

The bit rate of $>10$kbps is approximately more than two orders of magnitude higher than what can be achieved at a fiber distance of 20km without decoy states \cite{gobby2004b}. Indeed, if one were to employ more decoy states than the two we use here, there is not that much advantage to be gained. As shown in \cite{ma2005} the secure bit rate as a function of fiber distance for the single pulse and dual decoy pulse protocols varies significantly. However, the weak plus vacuum protocol is close to the asymptotic limit of using an infinite number of decoy states. 
This can be understood intuitively by the fact that the photon distribution is Poissonian and states with a photon number $N>2$ have a very small probability of occurring. Hence they will contribute little to the overall photon distribution when the average photon numbers of the signal and decoy states are $\mu$,$\nu<1$. Additionally, there are practical problems in using more decoy pulses such as a decrease of duty cycle of signal pulses, greater requirements on Alices' random number generator and more data processing power needed. 

The possibility of nonlinear effects being generated in the fiber due to the clock and signal pulses is negligible on the results of the quantum key distribution. As stated previously, both clock and signal pulses never overlap in time throughout the entire transmission rendering possible mixing effects negligible. Addtionally, any (small) Raman scattering generated by the clock pluse in the fiber was adequately filtered out using an efficient wavelength division multiplexer at Bob. In any case, if the Raman scattering was a problem, the quantum bit error rate would be higher than observed. Finally, the experimentally observed quantum transmittances agrees very well with the predicted transmittances indicating the system performs with the expected losses accurately.
\begin{figure}[htb!]
\centerline{\scalebox{1.25}{\rotatebox{0}{\includegraphics*[5mm,5mm][127mm,50mm]{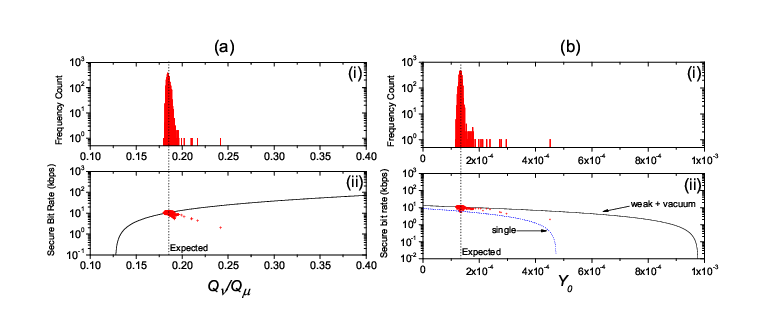}}}}
\caption{\label{fig:expected_values}(a)(i) Frequency count distribution of the quantum transmittance ratio $Q_{\nu}/Q_{\mu}$. (ii) The secure bit rate simulated as a function of $Q_{\nu}/Q_{\mu}$ (solid line); experimentally measured data (red crosses). (b)(i) Frequency count distribution of the vacuum count probability $Y_{0}$. (ii) The secure bit rate as a function of $Y_{0}$ (solid black line); experimentally measured data (red crosses). Also shown for comparison is the single decoy protocol secure bit rate (dotted line). The dotted lines in both Figs. show the expected values in the absence of PNS attacks, artifacts and manipulations of the vacuum count rates.}
\end{figure}

To assess the possibility of a PNS attack on the key distribution, the technique used previously \cite{yuan2007} was to monitor the ratio of transmittances $Q_{\nu}/Q_{\mu}$. If Eve decides to implement a PNS attack, the ratio $Q_{\nu}/Q_{\mu}$ will dip below the expected value as preferentially more multi-photon signals would be transmitted to Bob (the transmittance $Q_{\mu} > Q_{\nu}$ due to $\mu > \nu$). This is displayed in Fig. \ref{fig:expected_values}(a)(ii).  No secure bit rate is possible with $Q_{\nu}/Q_{\mu} < 0.13$.  

Now with an additional vacuum pulse at our disposal we can also check for Eve's manipulation of the vacuum rates. In keeping with the paranoid assumptions of GLLP \cite{gottesman2004}, we assume Eve can either reduce or increase the vacuum rates. Depicted in Fig. \ref{fig:expected_values}(b)(ii) is the simulated effect of changing $Y_{0}$ on the secure bit rate. For the weak plus vacuum protocol implemented here no secure bit rate is possible for $Y_{0}>10^{-3}$ (solid black line). However, if one were to use a single (non-vacuum) decoy protocol (dotted blue line) no secure bit rate would be possible for $Y_{0}>4.8\times10^{-4}$. This shows the power of using more than one decoy pulse. Further insight can be gained by examining the formulae for the weak plus vacuum protocol (eq.(\ref{eq:single_photon_gain}) \& eq.(\ref{eqn:error_rate})). The magnitudes of the single photon gain (privacy amplification) are greater (smaller) respectively by using the weak + vacuum protocol compared to employing the single pulse protocol. This is manifest through a tighter bound on the single photon gain eq. (\ref{eq:single_photon_gain}) and the single photon error rate eq. (\ref{eqn:error_rate}). The second term in eq. (\ref{eqn:error_rate}) reduces the overall single photon error rate due to the measurement of the vacuum pulses. In the single decoy pulse protocol this term is zero. 

Finally we address the issue of the small number of spikes observed in the vacuum counts transmittance (Fig. \ref{fig:transmittances_and_qbers}(a)). All classical messages are exchanged on the internet. We believe the spiking is due to increased internet traffic slowing down classical exchange of information between Alice and Bob and thereby affecting synchronization. It affected less than $1$\% of the entire set of quantum keys exchanged over the 60 hour period. For large numbers of counts (the signal and decoy pulses) this effect is negligible, but for low numbers of counts ($Y_{0}$) this effect can be apparent. We are currently working to improve this problem with modifications to the software. However, we note this behaviour does not compromise security as evident from Fig. \ref{fig:expected_values}(b)(ii). As can be seen, an increase in $Y_{0}$ overestimates the amount of privacy amplification and results in a shorter key, hence lower final secure key rate. The final secure key rate is underestimated for this small fraction of keys.

\section{Conclusions}
In summary, we have demonstrated practical one-way decoy pulse QKD over 20km of optical fiber. The key generation rate is $>10$ kbps on average over 60 hours. This is the highest reported key rate for this distance and duration. We believe the system to be a useful milestone in achieving a continuously operating QKD system. It is envisaged that such a system could be placed in a real-world environment such as a quantum nodal network with fiber links of around a few tens of kilometers of fiber. Such a system would be reliable and effective as a means of distributing quantum keys over a long period of time.
\section*{Acknowledgements}
The authors would like to thank the EC for funding under the FP6 Integrated Project SECOQC.


\end{document}